\documentclass[11pt,a4paper]{article}
\usepackage[cp1250]{inputenc}
\usepackage[T1]{fontenc} % w bilbiografii litera ą = {\k{a}}
\usepackage[font=small,skip=5pt]{caption}
\usepackage[english]{babel}
\usepackage{geometry}
\usepackage{graphicx}
\usepackage[iso]{isodateo}
\usepackage{floatflt}
\usepackage{algorithm}
\usepackage{algorithmicx} 
\usepackage{algpseudocode}
\usepackage{mathrsfs}		%dziwne literki Zbiorów Liczb
\usepackage{mathtools}
\usepackage{amsmath}
\usepackage{amsfonts}
\usepackage{amssymb}
\usepackage{algorithm}	%algorytmy
\usepackage{algorithmicx} %algorytmy
\usepackage{algpseudocode} %algorytmy
\usepackage{setspace}
\usepackage{xcolor}
\usepackage{multirow}
\usepackage{enumitem}

\DeclareMathAlphabet{\mathbbmsl}{U}{bbm}{m}{sl}
\newtheorem{mydef}{Definition} %definicja Definicji

\begin{document}

%{ \huge \bfseries A proposal for a new type of Petri nets with time parameters. }
\begin{center}
{ \huge \bfseries Extended time Petri nets}

\vspace{0.5cm}
Marcin Radom\,$^{1,2}$, Piotr Formanowicz\,$^{1}$
\end{center}
$^{1}$Institute of Computing Science, Poznan University of Technology, Poland\\
$^{2}$Institute of Bioorganic Chemistry, Polish Academy of Sciences

\vspace{20pt}
\section*{Abstract}
In many complex systems that can be modeled using Petri nets time can be a very important factor which should be taken into account during creation and analysis of the model. Time data can describe starting moments of some actions or their duration before their immediate effects start to influence some other areas of the modeled system. Places in a Petri net often describe static components of the system, but they can also describe states. Such a state can have time restrictions, for example, telling how long it can influence other elements in the model. Time values describing some system may be inconsistent or incomplete, which can cause problems during the creation of the model. In this paper, a new extension of time Petri nets is proposed, which allows the creation of models with different types of time data, which previously were possible to be properly used in separate types of well-known time Petri nets. The proposed new time Petri net solves this problem by integrating different aspects of already existing time Petri nets into one unified net. 

\vspace{7pt}
\textbf{Keywords:} Petri nets, time constraints, tokens with lifetimes

\section{Introduction}
Petri nets are one of the popular tools used to model complex systems. They can be used for systems characterized by a large number of elements and a high number of interactions between them. Since the introduction of Petri nets \cite{P62}, many modifications and extensions have been proposed. At a higher level of their classification, one can mention time nets, stochastic, continuous, hybrid, colored nets, and many more.

Time Petri nets (TPN) were first introduced in \cite{M74}. The first algorithmic approach to its analysis has been proposed in \cite{BM83}. In that paper, the so-called state classes have been defined and used to analyze the states of the net using state reachability graphs. This approach has been further improved in \cite{BD91}. In this type of Petri nets for each transition, a separate time interval is assigned. Each transition can fire only after a certain amount of time has passed since the moment of its activation. Active transitions in TPN must fire no later than the time defined as the second end of the range. In the same year as time Petri nets the concept of timed Petri nets, also called Petri (time) duration nets (DPN), was proposed \cite{R74}. Further analysis of this type of nets has been given \cite{S80}. In DPN, a certain measurable value is associated with each transition. In this type of time nets, each active transition is immediately fired, and tokens from its input places are taken at that moment. However, the production of new tokens occurs only after some time has passed, specified by the duration value assigned to the transition. An extension of this type of nets has been proposed (later called Interval-time Petri net, ITPN), where a fixed duration is replaced by a range \cite{PP12}. A separate category of time Petri nets are those in which values specifying time are assigned to places. These nets are called Petri nets with Time Windows \cite{P93}. In general, each token that appears in a given place has its own lifetime. If a place is described by a certain range, it is assumed that a token cannot stay in that place longer than the time specified by the upper value of this range. The lower value specifies the minimum amount of time that must elapse for a given token before it can be used to activate any transition. These three types of Petri nets with times are well known and widely used, and new extensions and proposals for their analysis are still being proposed, e.g., in \cite{ZFB+14, BBZ+16}.

Time is obviously an important factor in modeling and analysis of complex systems, for example biological ones. In such a system, new compounds have to form and mature, whereas the chemical reactions that are responsible for their creation are usually not instantaneous. One of the problems that can be encountered when modeling such a system using previously described types of time net is that the time data (such as time ranges, reaction duration, or lifetimes of system components) obtained from studies of biological systems can be multiple, incomplete or available in many forms. This may require additional calculations and changes for the time values obtained \cite{MLG+11, MGM17}. A new time Petri nets extension, with the features of the well-known time nets, would allow much more precise model development and adaptation for complex biological systems with many different time variables.
 
In this paper, a new type of Petri nets, called extended time Petri nets (xTPN) is proposed. It integrates three separate types of time nets that have been described, i.e., TPN with time intervals for an activation of transitions, ITPN which allows one to specify the duration of tokens production by transitions, and time nets in which the time ranges are associated with places. The use of three numerical time intervals, that is, two for transitions and one for places, can be convenient because it will allow both approximate and exact values of time to be included into the model. A brief introductory definition of a net of this type with two basic propositions for token analysis algorithms based on net time variables has been given in a conference paper \cite{RF22}. However, in that paper only some specific features of xTPNs, which could be potentially interesting from a biological modeling point of view, have been given. In this paper, we present full notation, definitions and semantics of the net. In the next sections, state changes and other interesting features, such as transformations, are given. It is worth mentioning that extended time Petri nets can be created and analysed using a recent version of software called Holmes \cite{SRF22} which is a program dedicated to the creation and analysis of different models using various types of Petri nets: classical, time, functional, etc.

The structure of this paper is as follows. At the beginning, definitions of more basic time Petri nets than xTPNS are given, and then a new proposed type of the nets is formally defined. A so-called $p$-state and $t$-state are introduced, and it is explained how these two substates together create a full state of an extended time Petri net. The activation of transitions depends on the existence of the so-called \emph{activating subsets of tokens}, which strongly depend on an time values associated with the net elements. This is crucial for understanding how the proposed net behaves and is explained in detail. In Section 3, some interesting features of the xTPN are given. First, a possible transformation of xTPNs elements (both places and transitions) is described. Next, two specific types of arcs, which are sometimes used in more basic types of Petri nets, i.e., read arcs and inhibitor arcs, are explained. They can behave in a different way in the proposed nets and the rules of such behavior are described. A summary of the new Petri time nets proposal is given at the end of the paper in Section 4. 

\section{Extended time Petri nets}\label{sec2}

The basic notation used in this paper is as follows. $\mathbb{N}$, $\mathbb{Q}$ and $\mathbb{R}$ denote sets of natural, rational, and real numbers, respectively. $\mathbb{N}^{+} = \mathbb{N} \setminus \{0\}$ denotes the set of natural numbers without 0, while $\mathbb{Q}^{+}_{0}$ denotes the set of non-negative rational numbers. On the other hand, $\mathcal{N}$, $\mathcal{W}$, $\mathcal{T}$, and $\mathcal{Z}$ denote various tuples representing different Petri nets.

A classical Petri net is the backbone of any proposed extension. In the literature it is sometimes called the \emph{skeleton}. It provides basic structural components, called places, transitions, and arcs. A classical Petri net is given in the following Definition \ref{def:PNwithoutState}.

\begin{mydef}\label{def:PNwithoutState} Petri net.\\
\textit{Petri net is a ordered 5-tuple $\mathcal{N} = (P, T, F, W, m_0)$, where:\\
$P$ and $T$ are finite, nonempty, and disjoint sets of places and transitions, respectively,\\
$F \subseteq (P \times T) \cup (T \times P)$ is a set of arcs,\\
$W: F \rightarrow \mathbb{N}^{+}$ is a function that assigns weights to the arcs,\\
$m_0: P\rightarrow \mathbb{N}$ is an initial marking \cite{M89}.
}
\end{mydef}

A structure of a Petri net is a bipartite directed graph, with vertices being places and transitions that are connected by arcs. Arcs have weights that are positive natural numbers. For example, in a model of some biological process, places can represent chemical substrates and products. In such a model, transitions will represent some elementary or more complex chemical processes. In places there are tokens, representing the number of elements of the system a given place represents. A state (also called marking) of a net is a vector that describes a number of tokens at each place at a given moment.

It is often convenient in the description of a Petri net to distinguish between a normal place and a transition and the so-called input/output \emph{ of} another place/transition. Formally, $p_i$ is an input place of transition $t_j$ if $(p_i, t_j) \in F$. Inversely, place $p_i$ is an output place of $t_j$ if $(t_j, p_i) \in F$. The set of input (output) places of transition $t_j$ is denoted as $^{\bullet}{t_j}$ (${t_j}^{\bullet}$). Similarly, input and output transitions relative to a given place $p_i$ can be defined, and their sets are denoted as $^{\bullet}{p_i}$ and ${p_i}^{\bullet}$, respectively. More precisely, we define four sets that have just been mentioned as follows:
\begin{description}
 \item[] $\quad\quad ^\bullet t = \{ p \in P : (p, t) \in F\}$ is a set of input places of $t$,
 \item[] $\quad\quad t ^\bullet = \{ p \in P : (t, p) \in F\}$ is a set of output places of $t$,
 \item[] $\quad\quad ^\bullet p = \{ t \in T : (t, p) \in F\}$ is a set of input transitions of $p$,
 \item[] $\quad\quad p ^\bullet = \{ t \in T : (p, t) \in F\}$ is a set of output transitions of $p$.
\end{description}

In addition to the above four sets, in a Petri net the so-called input/output places and transitions are often distinguished (i.e., without a reference to the other place or transition). For example, transition $t$ is an input transition in a Petri net if ${^\bullet}t = \emptyset$. Such a transition does not have any input places, and it is always active. On the other hand, transition $t$ for which $t{^\bullet} = \emptyset$ is called an output transition. When it fires, it consumes tokens from ${^\bullet}t$, but does not produce any tokens.

Transition $t$ in a Petri net becomes active if in each place $p_i \in {^\bullet}t$ there are at least as many tokens as the value of weight $W(p_i, t)$. An active transition can fire, i.e., it produces tokens in its output places, yet it is not always assumed that it immediately must fire when it becomes active. If, however, a transition fires, it produces tokens in each place $p_j \in t{^\bullet}$ in quantities equal to weight $W(t, p_j)$, while at the same time it takes tokens from each $p_i \in {^\bullet}t$ in quantities equal to weight $W(p_i, t)$.

There are many variants of Petri nets with the ability to handle time values assigned to transitions. Two especially well-known nets have been introduced in \cite{M74} (Time Petri net (TPN) proposed by Merlin in 1974) and in \cite{R74} (Timed Petri net (DPN) also known as Duration-time Petri net proposed by Ramchandani also in 1974). In Definition \ref{def:TPNmerlin} the first of them is given.

\begin{mydef}\label{def:TPNmerlin} Time Petri net.\\
\textit{A Time Petri (TPN) net is an ordered 6-tuple $\mathcal{T} = (P, T, F, W, m_0, I)$ such that:\\
the 5-tuple $S(\mathcal{T}) = (P, T, F, W, m_0)$ is a Petri net,\\
$I : T \rightarrow {\mathbbmsl{Q}}_{0}^{+} \times ({\mathbbmsl{Q}}_{0}^{+} \cup \{+\infty\}) $ and for each $t \in T$, with $I(t) = (I_1(t), I_2(t))$ it holds that $I_1(t) \leq I_2(t)$. \cite{P13}
}
\end{mydef}

The classical Petri net denoted here as 5-tuple $S(\mathcal{T})$ is called a \emph{skeleton} of $\mathcal{T}$. $I$ is interval function, for each transition $t$ returning values $I_1(t)$ and $I_2(t)$ being rational numbers, which are called the earliest and the latest firing times of $t$, respectively. An active transition will fire only after some time $\tau^{act}_t$ for which it holds $I_1(t) \leq \tau^{act}_t \leq I_2(t)$.

The timed Petri net (DPN) proposed by Ramchandani in \cite{R74} has a different time function, which in that case for each transition assigns a single value. This deterministic value describes a production time for a given transition. In this paper, we will use an extension of DPN, called an interval-timed Petri net \cite{PP12, P13}, in which instead of a single deterministic value, another range is assigned. This kind of net is given in Definition \ref{def:ITPNpopova}.

\begin{mydef}\label{def:ITPNpopova} Interval-timed Petri net.\\
\textit{Interval timed Petri net (ITPN) is an ordered 6-tuple $\mathcal{W} = (P, T, F, W, m_0, D)$ such that:\\
the 5-tuple $S(\mathcal{W}) = (P, T, F, W, m_0)$ is a Petri net,\\
$D : T \rightarrow {\mathbbmsl{Q}}_{0}^{+} \times {\mathbbmsl{Q}}_{0}^{+}$ with $D(t) = (D_1(t), D_2(t))$ implies $D_1(t) \leq D_2(t)$ for every $t \in T$. \cite{P13}
}
\end{mydef}

In this case, interval function $D$ assigns two values for each transition, but they have different meanings from the values of $I$ in Definition \ref{def:TPNmerlin}. We call $D_1(t)$ the \emph{earliest firing duration} and $D_2(t)$ the \emph{latest firing duration}. However, the assumed changes in such net behavior are slightly more significant than can be seen in the definition of it. In a classical Petri net, it is often assumed that an active transition can fire, but it is not required to immediately do so. In TPN, an active transition remains in an activation state for a time $\tau^{act}_t$, then it fires. Firing itself remains an immediate event that, at the same time, takes the necessary tokens from all input places and produces new ones in the output places of $t$. In both, the DPN proposed by Ramchandani and the interval-time Petri net $\mathcal{W}$, there are assumed changes in this mechanism. First, token production takes time, defined by a single value, such as in the DPN, or by a time within a range $[D_1(t), D_2(t) ]$. Second, it is assumed that each active transition must fire immediately and take necessary amount of tokens from its input places. If one would allow the behavior known from a classical Petri net in which an active transition may or may not fire, this would cause problems in determining exact times of transition firing sequences. Therefore, it is assumed that every transition that has enough tokens in its input places to be active immediately fires. It is often called \emph{a maximum firing rule}.

When considering $\mathcal{W}$ from Definition \ref{def:ITPNpopova}, firing works as follows. Immediately when transition $t$ becomes active, it starts its firing event. At that moment, tokens are taken from the input places. Then, an internal timer for a transition starts counting the time towards some value $\tau^{prod}_t$ for which $D_1(t) \leq \tau^{prod}_t \leq D_2(t)$. Exactly after time $\tau^{prod}_t$, transition $t$ produces tokens at its output places.

In this paper, a new type of time Petri net is proposed which has some properties of the described nets $\mathcal{T}$ and $\mathcal{W}$, and also uses the idea of time associated with places. Petri nets in which time is assigned to places are not as popular as the ones discussed here, and their notation varies depending on the task for which they are used. The extended time Petri net proposed in this paper has its own notation for such places, while the general idea was proposed in \cite{P93} and later analyzed in \cite{WP09}. They are an extension of the nets introduced in \cite{JR83}. It is important to mention one other extension of timed nets. In \cite{BLT90}, a so-called Petri net in time arcs has been proposed. In such nets, a time interval is associated with arcs, which then allow tokens to pass only if their lifetime is within the range. Some interesting properties of such nets have been analyzed in \cite{JJS11}. 
%In the proposed extended time Petri-nets we will use a notation where time interval is assigned to places.

Now we can proceed to the basic definitions of an extended time Petri net. In the proposed net, function $I$ assigns four rational numbers to each transition, which are, in fact, two time intervals. The first two values cover the functionality of the original function $I$ of $\mathcal{T}$. The second interval determines the duration of token production. Finally, we introduce the function $J$ that assigns time constraints for places. An extended time Petri net without a state is described in Definition \ref{def:xTPNstateless}.

\begin{mydef}\label{def:xTPNstateless} Extended time Petri net without state.\\
An extended time Petri net without state is an ordered 3-tuple $\mathcal{Z} = (N, I, J)$, where:\\
$N = (P, T, F, W)$ is a classical Petri net,\\
$I : T \rightarrow {\mathbbmsl{Q}}_{0}^{+} \times ({\mathbbmsl{Q}}_{0}^{+} \cup \{+\infty\}) \times {\mathbbmsl{Q}}_{0}^{+} \times ({\mathbbmsl{Q}}_{0}^{+} \cup \{+\infty\})$ and for every $t \in T$, where $I(t) = (I_1(t), I_2(t), I_3(t), I_4(t))$ it holds that $I_1(t) \leq I_2(t)$ and $I_3(t) \leq I_4(t)$,\\
$J : P \rightarrow {\mathbbmsl{Q}}_{0}^{+} \times ({\mathbbmsl{Q}}^{+} \cup \{+\infty\})$ and for every $p \in P$, where $J(p) = (J_1(p), J_2(p))$ it holds that $J_1(p) < J_2(p)$.
\end{mydef}

Function $I$ assigns to each transition four ordered values $I_1(t), I_2(t), I_3(t), I_4(t)$, which are in fact two pairs defining some specific time intervals. The first pair of values, i.e., $I_1(t)$ and $I_2(t)$ specify the minimum and maximum time after and before which an active transition has to be fired. Later in the text, they will often be referred to as $\alpha$ values, where $I_1(t_i) = \alpha^{L}_{t_i}$ and $I_2(t_i) = \alpha^{U}_{t_i}$. The letters $L$ and $U$ refer to the lower and upper bounds of a range, respectively. These two values always satisfy the inequality $\alpha^{L}_{t_i} \leq \alpha^{U}_{t_i}$ for any transition $t_i \in T$. 

The second pair of values $I_3(t)$ and $I_4(t)$ specifies the minimal and maximal token production time for a transition that has been fired. They will be later referred to as $\beta$ values: $I_3(t_i) = \beta^{L}_{t_i}$ and $I_4(t_i) = \beta^{U}_{t_i}$ for each transition $t_i \in T$.

There are a few features of such transitions which must be explained. First of all, the extended time Petri net transitions do not need to immediately start tokens production after their activation. The transition starts the production of tokens after some time $\tau^{\alpha}_{t}$. It is not exactly known how long activation will last, however, it is always true for $\tau^{\alpha}_{t}$ that $0 \leq \alpha^{L}_t \leq \tau^{\alpha}_{t} \leq \alpha^{U}_t$. After that time, tokens are consumed from input places. The time of production is not strictly deterministic (i.e., it is not a single value like in DPN), but is described by a closed interval specifying a minimum and maximum time of token production (this  is a feature of ITPN transitions). When production starts, transition immediately takes tokens from its input places, but the production of new tokens in the output places occurs only when time $\tau^{\beta}_{t}$ has elapsed. Exactly how long it will take is unknown, however, it is always true that $0 \leq \beta^{L}_t \leq \tau^{\beta}_{t} \leq \beta^{U}_t$. It should be noted that over time $\tau^{\alpha}_{t}$ a transition must remain active. On the other hand, when token production has started, the conditions for transition activation are not checked because they are not relevant.

It will be formally defined later that each transition is at any moment in one of the three states: inactive, active, or producing tokens. With each transition two timers are associated. The first, later referred to as $u_t$ is the activation timer. When time $\tau^{\alpha}_{t}$ has passed, the transition will start producing tokens. The production state begins with tokens being taken from $^{\bullet}t$. When production starts, the second timer $w_t$ starts counting. New tokens will be created after time $\tau^{\beta}_{t}$. The inactive transition has its timers set to a special symbol $\#$. A more detailed description will be given in the section describing a so-called $t$-state of $\mathcal{Z}$, in Definition \ref{def:xTPNtstate}.

\begin{figure}[ht]
\centering
    \includegraphics[width=0.5\textwidth]{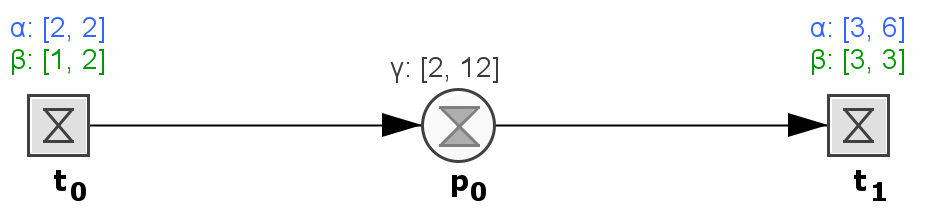}
    \caption{Extended time Petri net example with time intervals $\alpha$ and $\beta$ assigned to transitions and time interval $\gamma$ assigned to place. Transition of the extended time Petri net are represented by rectangles with hourglass, place is represented by a circle. It should be noted that both input and output transition can have two intervals assigned, even though, e.g., $t_0$ will not consume any tokens when in fires, nor $t_1$ will produce any tokens. Production time for $t_1$ is deterministic, because both $\beta$ values for $t_1$ are equal. The same is true for an activation time of $t_0$, that is $\tau^{\alpha}_{t_0} = 2$.}
		\label{fig:1}
\end{figure}

Function $J$ assigns to each place a pair of numbers $J_1(p)$ and $J_2(p)$ specifying the closed time interval for a place. The lower value denotes the minimum maturity time for a token, before which it cannot be used for activation or production. The higher value specify the maximum time a token can stay at that place. These values will be referred to as $\gamma$ values such that $J_1(p_j) = \gamma^{L}_{p_j}$ and $J_2(p_j) = \gamma^{U}_{p_j}$. Although $\gamma^{L}_{p_j}$ can be zero, $\gamma^{U}_{p_j}$ must be a positive number (i.e., the maximum token lifetime allowed in a place must be longer than $0$). The value of $\gamma^{L}_p$ specifies the time that must elapse separately for each token before it can activate output transitions of a place, where it resides. A total lifetime of a token in a given place cannot exceed $\gamma^{U}_p$. When the age of a token is equal to $\gamma^{U}_p$, it has to be used by some output transition or it is removed from the given place. A simple example of such a net is presented in Figure \ref{fig:1}.

Since tokens are separate objects with individual lifetimes, a special type of set must be defined to store values representing such objects. A special class of $K$ multisets is given in Definition \ref{def:xTPNmultisetK}. Such multisets are associated with places and contain values that represent tokens of a given place.

\begin{mydef}\label{def:xTPNmultisetK} Multiset type K.\\
Let $\prescript{a}{b}{K}$, where $a > 0$ and $0 \leq b \leq a$, be a multiset of non-negative numbers less than or equal to some positive value $a$. Numbers of such a multiset must have values belonging to the closed interval $[b, a]$. In other words, $\kappa_x$ belongs to $\prescript{a}{b}{K}$ only if $\kappa_x \in {\mathbbmsl{R}}_{0}^{+}$ and $0 \leq b \leq \kappa_x \leq a$. A special case is a multiset $\prescript{a}{b}{K} = \{\#\}$.
\end{mydef}

In the above definition, the value $a$ is an admissible maximum for the elements of $\prescript{a}{b}{K}$. The value $b$ is the minimum allowed value of the elements of the multiset $K$. The multisets $\prescript{a}{b}{K}$ will be assigned to the places or will describe in some other way the distribution of tokens in $\mathcal{Z}$. For example, multisets $K$ assigned to places to describe their tokens will always have value $b = 0$. On the other hand, for example, activating subsets of tokens which are also multisets of type $K$, will usually have a non-zero value of $b$. As a general rule, the index on the upper right of the multiset $K$ will always inform about the specific functions of that multiset. A multiset of this kind can be empty, that is, $K = \emptyset$. The case where $K = \{\#\}$ will be used later, for example, in definitions dealing with transition activation constraints.

After assigning to a specific place, the name of that place will be given as the bottom right index of the multiset $K$, for example, as $\prescript{a_j}{b_j}{K}_{p_j}$. Since a positive non-zero value $\gamma^{U}_{p_j}$ is associated with each place $p_j$, assigning a multiset $\prescript{a_j}{b_j}{K}$ to a particular place $p_j$ will precisely set the values $a_j$ and $b_j$ such that $a_j =\gamma^{U}_{p_j}$ and $b_j = 0$. Values $\gamma^{L}_{p_j}$ different from zero will be used for some other specific types of such multisets. 
%If for some reason the exact values $a_j$ or $b_j$ are not important at a given time, we will simply name such a multiset describing tokens of $p_j$ as $K_{p_j}$. 

In the next Definition \ref{def:xTPNmultisetM} \emph{multiset of multisets}, a special is given.

\begin{mydef}\label{def:xTPNmultisetM} Multiset type M.\\
Let $M$ be a multiset of size $n$ whose elements are multisets $K$. Therefore, $M = \{\prescript{a_1}{b_1}{K}_1, \prescript{a_2}{b_2}{K}_2, ..., \prescript{a_n}{b_n}{K}_n \}$ and $n = |P|$.
\end{mydef}

This type of multisets generally describes a so-called $p$-state of the net, which contains data about the lifetimes of all tokens that exist at different places in a given moment. Multisets $M$ will have some specific subtypes defined later, but will always satisfy $|M| = |P|$. 

In Table \ref{tab:symbols} currently discussed symbols are presented together with the ones that will be explained in the next section of the paper. This table my serve as a reference to used symbols and notation, each pointing to a specific definition given in the paper.

\begin{table}[ht]
	\begin{center}
		\caption{Symbols  definitions}
		 \label{tab:symbols} 
		\begin{tabular}{ |p{3cm}|p{12.5cm}|  }
		 %\hline
		 %\multicolumn{3}{|c|}{Possible transformations} \\
		 \hline
		 Symbols & Notation meaning with a reference to a specific Definition in the paper \\
		 \hline \hline
	   [$\alpha^{L}_{t}$,  $\alpha^{U}_{t}$], [$\beta^{L}_{t}$, $\beta^{U}_{t}$] & Activation and production intervals for transition, respectively. Explained in short discussion directly after Definition \ref{def:xTPNstateless}. \\ \hline
	$\gamma^{L}_{p}$,  $\gamma^{U}_{p}$ & Token maturity and time limit (i.e., maximum lifetime), explained in discussion after Definition \ref{def:xTPNstateless}. \\ \hline
		$\prescript{a}{b}{K}_{p}$ & Multisets type K containing numbers describing tokens in place $p$. There are a few subtypes of them used. Their most general description given in Definition \ref{def:xTPNmultisetK}. \\ \hline
    $M$ & Multiset type $M$, always containing multisets $K$ for all places, therefore always $|M| = |P|$. Introduced in Definition \ref{def:xTPNmultisetM}. \\ \hline
        $m(p) = \prescript{\gamma^{U}_{p}}{0}{K}_{p}$  & Function defining $p$-state (Definition \ref{def:xTPNpstate}), assigns multiset representing tokens to every net place.\\ \hline
    $h(t) = (u_t, w_t)$ & Function defining $t$-state (Definition \ref{def:xTPNtstate}), timer $u_t$ counts activation time, timer $w_t$ counts production time.\\ \hline
        $M' \subset M$, $\quad\quad\quad$ $M' \not\subset M$ & Inclusion of multisets, introduced in Definition \ref{def:xTPNkinclusion}. Notation used in transition activation, explained in Definition \ref{def:xTPNtransitionActive}. \\ \hline
    $\prescript{a_j}{0}{K}_{p_j} \xrightarrow{\tau} \prescript{a_j}{0}{K'}_{p_j}$, 
        $M \xrightarrow{\tau} M'$ & Elapsing of time $\tau$ in multisets, given in Definition \ref{def:xTPNtimeLapse}. \\ \hline
    $\prescript{a}{b}{K}_{p} + v$, $\prescript{a}{b}{K}_{p} - v$ & Notation explained in Definition \ref{def:xTPNaddTokens} and \ref{def:xTPNremoveTokens}, respectively, used to denote adding or removing $v$ tokens from a place. \\ \hline
        $M_1 \:\langle\: M_2$, $\quad\quad\quad$ $M_1 \:\rangle\: M_2$ & M-addition and M-subtraction of two multisets type $M$, given in Definition \ref{def:xTPNsetMaddition} and \ref{def:xTPNsetMsubtraction}, respectively.  \\ \hline
    $M^{t \oplus}$, $M^{t \ominus}$ & Two special multisets describing tokens added to places ($M^{t \oplus}$) when transition $t$ ends its production or removed ($M^{t \ominus}$) when $t$ starts production phase. Given in Definition \ref{def:xTPN_K_plus_set} and \ref{def:xTPN_K_minus_set}, respectively. \\ \hline
		\end{tabular}
	\end{center}
\end{table}

\subsection {State of extended time Petri net}

A state of extended time Petri net $\mathcal{Z}$ consists of two substates. They will be called $p$-state and $t$-state, respectively, which describe the token distribution at places and time values assigned to transitions. Each token of the proposed net is a distinguishable object with finite lifetime. Even when two tokens in one place have identical time values, they are considered to be independent, different objects in the net. Definition \ref{def:xTPNpstate} describes the $p$-state of the net.

%\begin{mydef}\label{def:xTPNpstate} p-state of extended time Petri net $\mathcal{Z}$.\\
%Let $P$ be the set of all places in $\mathcal{Z}$, where $n = \vert P\vert$, multiset $M' = \{\prescript{a'_1}{0}{K'}_1, \prescript{a'_2}{0}{K'}_2, ..., \prescript{a'_n}{0}{K'}_n\}$ and for each place $p_j$ value $\gamma^{U}_{p_j} = a_j$. The p-state of $\mathcal{Z}$ is a function $m : P \rightarrow M$ assigning elements of the multiset $M = \{\prescript{a_1}{0}{K}_1, \prescript{a_2}{0}{K}_2, ..., \prescript{a_n}{0}{K}_n\}$ to places such that $m(p_j) = \prescript{a_j}{0}{K}_{j}$. Each multiset $K_{j} \in M$ is formed from the respective multiset $K'_{j} \in M'$ in such a way that: $\prescript{a_j}{0}{K}_{j} = \{\kappa_x \in \prescript{a'_j}{0}{K'}_j \land \kappa_x \leq a_j\}$.
%\end{mydef}

%It should be noted that the function $m$ assigns multisets $\prescript{a_j}{0}{K}_{j}$ which are subsets of multisets $\prescript{a'_j}{0}{K'_j}$ from $M'$. $M'$ can be any multiset of size $n$, whose elementary multisets $K'$ have elements greater than or equal to zero. However, a proper multiset $\prescript{a_j}{0}{K}_{j}$ assigned to a specific place $p_j$ in $\mathcal{Z}$ must have $a_j = \gamma^{U}_{p_j}$. Therefore, the multiset $\prescript{a_j}{0}{K_j}$ assigned to the place $p_j$ has only the corresponding elements of $\prescript{a'_j}{0}{K'_j}$ in $M'$ that are equal to or less than $a_j = \gamma^{U}_{p_j}$. 

\begin{mydef}\label{def:xTPNpmarking} p-marking of extended time Petri net $\mathcal{Z}$.\\
Let $P$ be the set of all places in $\mathcal{Z}$, where $n = \vert P\vert$, multiset $M = \{\prescript{a_1}{0}{K}_1, \prescript{a_2}{0}{K}_2, ..., \prescript{a_n}{0}{K}_n\}$. The p-state of $\mathcal{Z}$ is a function $m : P \rightarrow M$ assigning elements of the multiset $M$ to places such that $m(p_j) = \prescript{a_j}{0}{K}_{j}$. 
\end{mydef}

\begin{mydef}\label{def:xTPNpstate} p-state of extended time Petri net $\mathcal{Z}$.\\
Let $P$ be the set of all places in $\mathcal{Z}$, where $n = \vert P\vert$ and m is a p-marking of $\mathcal{Z}$. Function m is p-state of $\mathcal{Z}$ if for each multiset $\prescript{a}{0}{K}_j \in M$ assigned to place $p_j$ it holds that $a = \gamma^{U}_{p_j}$.
\end{mydef}

Evidently, not every p-marking is also a proper p-state. For that it is required, that each element of assigned multiset $\prescript{a}{0}{K}_j$ is less or equal than $\gamma^{U}_{p_j}$.

%It should also be noted that the value $\gamma^{L}_{p_j}$ is not considered when determining which elements of $\prescript{a'_j}{0}{K'}_j$ are allowed in multiset $\prescript{a_j}{0}{K}_{j}$ that is then assigned to the place $p_j$. A nonzero positive value of $\gamma^{L}_{p_j}$ will be used in some other subtypes of such multisets.

Given a $p$-state of the net (i.e., the distribution of tokens: their number and their time values in all places), it is possible to determine whether a transition is active (denoted as $t^{act}$), inactive ($t^{\#}$) or is producing tokens ($t^{prod}$). These are the three basic states of any transition in $\mathcal{Z}$.  Definition \ref{def:xTPNtstate} introduces the $t$-state of $\mathcal{Z}$.

\begin{mydef}\label{def:xTPNtstate} t-state of $\mathcal{Z}$.\\
Let $T$ be the set of all transitions of $\mathcal{Z}$. Function $h$ describes the t-state of $\mathcal{Z}$ and is defined as $h : T \rightarrow ({\mathbb{R}}_0^{+} \cup \{\#\}) \times ({\mathbb{R}}_0^{+} \cup \{\#\})$. Function $h$ assigns to each transition an ordered pair of elements $(u_t, w_t)$ in such a way that at most one element of a pair can be a real number. If for a given transition $u_t \in {\mathbb{R}}_0^{+}$, then $w_t = \#$, and if $w_t \in {\mathbb{R}}_0^{+}$, then $u_t = \#$. Additionally, when an element of a pair is a real number, we have $0 \leq u_t \leq \alpha^{U}_t$ or $0 \leq w_t \leq \beta^{U}_t$. A special case is where $h(t) = (\#, \#)$.
\end{mydef}

Transition $t$ can be in one of three $t$-states given by function $h$, depending on the values of the elements in the pair $(u_t, w_t)$. The possible three cases are as follows:
\begin{enumerate}
 \item $h(t) = (u_t = \#, w_t = \#)$ specifies that transition $t$ is inactive and it does not produce tokens. This state is denoted as $t^{\#}$.
 \item $h(t) = (u_t \in {\mathbb{R}}_0^{+}, w_t = \#)$ specifies that transition $t$ is currently active and does not produce tokens. This state is denoted as $t^{act}$.
 \item $h(t) = (u_t = \#, w_t \in {\mathbb{R}}_0^{+})$ describes a transition currently in production state. It should be noted that in this state activation rules (as explained later) do not apply. This state is denoted as $t^{prod}$.
\end{enumerate}

A transition that is inactive can become active, which will be formally described later. A transition in the active state can either stop being active or start the production of tokens. It is important to discuss the third case, when a transition is in the production state. 
%This transition is denoted as $t^{prod}$ and is described by a pair: $(u_t = \#, w_t \in {\mathbb{R}}_0^{+})$. 
%When this state begins, the transition takes tokens from its input places and remains in this state for all time $\tau^{\beta}_{t}$. 
In this state its potential activation or deactivation is not considered. The activation state can be stopped before the start of a production state. However, the production state will always last for $\tau^{\beta}_{t}$ time units, even if the conditions for the transition to be active are no longer satisfied, because they are simply not considered during that time. When production ends, the transition changes its $t$-state to active or inactive.

Now the full state of extended time Petri net can be given in Definition \ref{def:xTPNzstate}.

\begin{mydef}\label{def:xTPNzstate} State of extended time Petri net $\mathcal{Z}$.\\
%Let the extended time Petri net $\mathcal{Z}$ be given, let function $m$ determine its $p$-state and function $h$ determine its t-state. 
The state of extended time Petri net $\mathcal{Z}$ is the pair $z = (m, h)$.
\end{mydef}

\subsection{Transition activation}\label{sec3}
In order to determine if a given transition $t$ is active, one must look more closely at the elements of all multisets $K_{p_j}$, for $p_j \in {^{\bullet}t}$. Suppose, for example, that place $p_j$ is the only input place of transition $t_i$ and they are connected by an arc with weight equal to $2$. For transition $t_i$ to be active the multiset $K_{p_j}$, that describes tokens of place $p_j$, must contain at least 2 numbers $\kappa_1$ and $\kappa_2$ such that $\kappa_1 \geq \gamma^{L}_{p_j} \land \kappa_2 \geq \gamma^{L}_{p_j}$.

To formally determine what it means for a transition to be active, a special subset of multiset $K_{p_j}$ must be defined which will be called an \emph{activating subset of tokens}. This subset is given in Definition \ref{def:xTPNactSubset}. 
%For simplicity of notation, we will omit the value $a_j = \gamma^{U}_{p_j}$ in $\prescript{a_j}{b_j}{K}_{p_j}$ for a given place $p_j$, which in that definition is not relevant.

\begin{mydef}\label{def:xTPNactSubset} Activating subset of tokens.\\
Let $\prescript{a_j}{0}{K}_{p_j}$ be the multiset of tokens of place $p_j$, $t$ is output transition for $p_j$, $w \in \mathbb{Z}^{+}$ be the weight of arc $W(p_j, t)$, $a_j = \gamma^{U}_{p_j}$ and $b_j = \gamma^{L}_{p_j}$. Let $\prescript{a_j}{b_j}{K}_{p_j}^{sub} \subseteq \prescript{a_j}{0}{K}_{p_j}$ and $\prescript{a_j}{b_j}{K}_{p_j}^{sub} = \{\kappa \in \prescript{a_j}{0}{K}_{p_j} : \kappa \geq b_j\}$. If $\vert\prescript{a_j}{b_j}{K}_{p_j}^{sub}\vert \geq w$, then we call this subset an activating subset of tokens in $p_j$ for $t$ and denote it as $\prescript{a_j}{b_j}{K}_{p_j-t}^{act}$.
\end{mydef}

Transition activation occurs only, if for each place $p_j \in {^\bullet}t$ there exists an activating subset of tokens. In simpler words, one can say that in order for a transition to be active, there must be a sufficient number of tokens (determined by an arc weight) in all its input places. The tokens that are considered when creating that activating subset are those with a sufficiently long life in $p_j$ (given by the value $b_j = \gamma^{L}_{p_j}$).

It should be noted that if transition $t$ has many input places, for example $p_1$ and $p_2$, in order for it to be active, it must have activating subsets in both $K_{p_1}$ and $K_{p_2}$. To distinguish between them, the lower right indexes of such activating subsets will be given in a form place-transition, that is, in that example there must exist $\prescript{a_1}{b_1}{K}_{p_1-t}^{act} \in \prescript{a_1}{0}{K}_{p_1}$ and $\prescript{a_2}{b_2}{K}_{p_2-t}^{act} \in \prescript{a_2}{0}{K}_{p_2}$. 
%For simplicity, value $a_j$ will be omitted later, since for both multisets (i.e., the one describing a place and its potential activating subset if it exists) it is always the same.

One important definition is required to adequately describe whether a given transition is active in state $z$. In Definition \ref{def:xTPNkinclusion} an inclusion relation of two $M$ multisets is given. Later, it will be used, for example, to determine if a given transition is active in the current state of $\mathcal{Z}$. 

\begin{mydef}\label{def:xTPNkinclusion} Inclusion of multisets $M$.\\
For two multisets of the same size $M = \{ K_1, \cdots, K_n\}$ and $M' = \{K'_1, \cdots, K'_n\}$ we say that $M'$ is included in $M$, which is formally written as $M' \subset M$ if $\displaystyle\mathop{\forall}_{j=1, ..., n}  K'_j \subseteq K_j$. Otherwise, $M' \not\subset M$.
\end{mydef}

Definition \ref{def:xTPNtransitionActive} formally describes the conditions for transition activation.

\begin{mydef}\label{def:xTPNtransitionActive} Transition activation.\\
Let $t \in T$ be a transition in $\mathcal{Z}$ that in a given state $z$ does not produce tokens. Let $M$ be a multiset which elements are assigned to places by function m. %${^{\bullet} t} \in P$ be a subset of input places of $t$, $n = |P|$, $a_j = \gamma^{U}_{p_j}$ and $b_j = \gamma^{L}_{p_j}$. 
To check whether $t$ can be active in state $z$, multiset $M^t =\{K_{p_1-t}, \cdots, K_{p_n-t}\}$ is created using elements of $M$, such that each multiset $K_{p_j-t} \in M^t$ are defined as follows:
\[
K_{p_j-t} = \begin{cases}
\prescript{a_j}{b_j}{K}_{p_j-t}^{act} \quad \:\: \textit{if} \:\: p_j \in {^{\bullet} t} \land |\prescript{a_j}{b_j}{K}_{p_j-t}^{sub}| \geq W(p_j, t)\\
\{\#\} \quad\quad\quad \textit{if} \:\: p_j \in {^{\bullet} t} \land |\prescript{a_j}{b_j}{K}_{p_j-t}^{sub}| < W(p_j, t)\\
\varnothing \quad\quad\quad\quad \textit{if} \:\: p_j \notin {^{\bullet} t}\\
\end{cases}.
\]
A transition $t$ is active in the p-state described by $M$ if $M^t \subset M$.
\end{mydef}

The above definition uses the notation introduced in Definitions \ref{def:xTPNactSubset} and \ref{def:xTPNkinclusion}. The former defines an activating subset of tokens with a sufficient lifetimes. Multiset $M = \{\prescript{a_1}{0}{K}_1, \prescript{a_2}{0}{K}_2, ..., \prescript{a_n}{0}{K}_n\}$ for $n = |P|$ describes tokens in all net places. If $p_j \in {^{\bullet} t}$ and its corresponding multiset $\prescript{a_1}{0}{K}_j$ contains activating subset of tokens (Definition \ref{def:xTPNactSubset}), then it will also be an element of $M^t$. If activating subset of tokens for $p_j$ does not exist, then $K_{p_j-t} = \{\#\}$. 

\begin{figure}[ht]
\centering
    \includegraphics[width=0.6\textwidth]{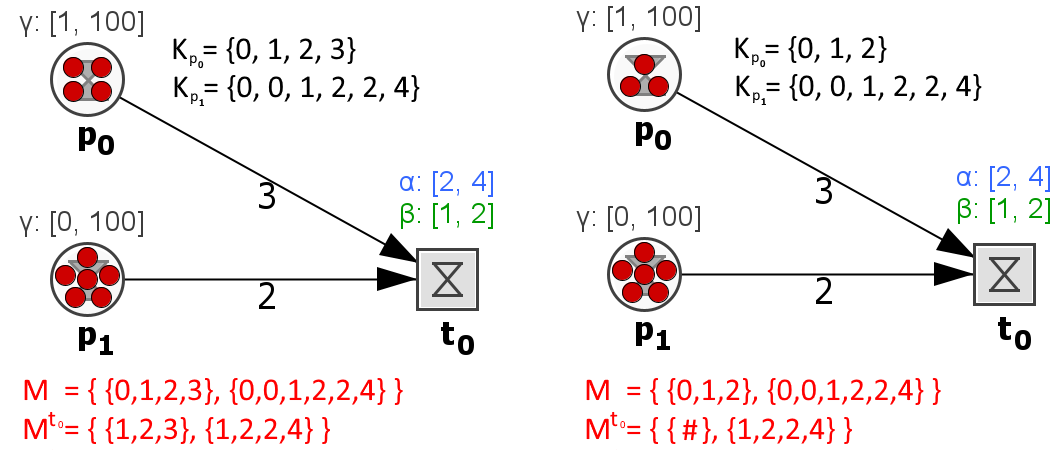}
    \caption{On the left side example transition $t_0$ is active, because in both $p_0$ and $p_1$ there is enough number of old enough tokens to form activating subsets. On the right side example transition $t_0$ is not active, because in $p_0$ there are only two old enough tokens in $K_{p_0}$ (represented by numbers 1 and 2). Therefore, multiset $M^{t_0}$ representing activating subsets for $t_0$ contains $\{\#\}$ instead of a proper activating subset for $p_0$. As a result, $M^{t_0}$ in right side example is not a subset of $M$.}
		\label{fig:2}
\end{figure}

In a case where $K_{p_j} = \{\#\}$ for any $p_j \in {^{\bullet} t}$, then the set $M^t$ will not be a subset of $M$ (Definition \ref{def:xTPNkinclusion}), which means that the transition $t$ is inactive. Figure \ref{fig:2} shows an example of two transitions: active and inactive according to the rules described in Definition \ref{def:xTPNtransitionActive}.

\begin{figure}[ht]
\centering
    \includegraphics[width=0.7\textwidth]{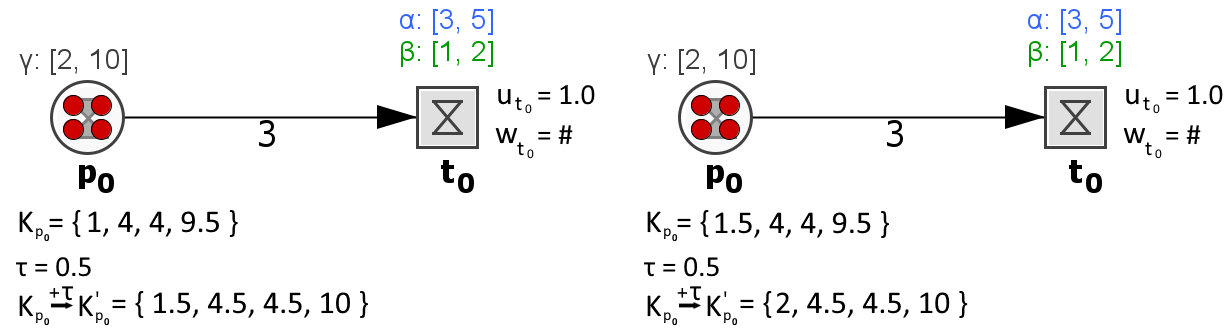}
    \caption{In both left and right cases it is assumed that time progress by $\tau = 0.5$ units. The only, yet very important difference between these cases is, that on the left the youngest token have a lifetime equal to 1, while on the right side its lifetime is equal to 1.5. This will result in deactivation of $t_0$ after time greater than $\tau$ by any measurable value, while in the example on the right $t_0$ will remain active.}
		\label{fig:3}
\end{figure}

Another example in Figure \ref{fig:3} shows how time influences an activation of the transition. In this figure a notation for updating time in $K$ multisets is used at the bottom: $\prescript{a}{b}{K}_{j} \xrightarrow{+\tau} \prescript{a}{b}{K'}_{j}$. It is formally explained later and in general it means that all elements of $K$ have been increased by $\tau$.

The example on the left-hand side of Figure \ref{fig:3} shows a situation in which, after a time longer than $\tau = 0.5$, transition $t_0$ will stop being active. This comes from the fact that the oldest token in $p_0$ represented by the value 9.5 can remain in that place for no longer than $\tau$ (its lifetime cannot exceed the value $\gamma^{U}_{p_0} = 10$). After \emph{exactly} that time, it can be seen that the multiset $K'_{p_0}$ still has enough tokens to support activation. However, transition $t_0$ cannot fire yet, because $u_{t_0} + \tau < \alpha^{L}_{t_0}$. After a time slightly longer than $\tau = 0.5$, for example, $\tau'$ such that $\tau' + \kappa_4 > \gamma^{U}_{p_0}$, this oldest token will be removed. Three tokens will remain in $p_0$, however, the youngest will still have lifetime less than $\gamma^{L}_{p_0}$. Since three tokens with sufficient lifetimes are required to form activating subset and only two remained, $t_0$ in this left example will stop being active.

As can be seen in the example on the right side, in that case, after time $\tau = 0.5$ the youngest token has already reached a minimum age to support activation of $t_0$ (that is, multiset $K'_{p_0} = \{2, 4.5, 4.5, 10\}$). Even when the oldest one will be removed, three remaining still form an activating subset. This example also shows that the tokens in the activating subset can change over time, but the activation subset remains.

\subsection{Operations on multisets of type $K$ and $M$}

This section will define possible operations on multisets $K$, specifying the number and current ''age'' of tokens in places. In this section, some specific sub-types of multisets $K$ and $M$ will also be introduced. In particular, the following operations should be precisely defined:

\begin{itemize}
	\item Elapsing of time - it means increasing the values of elements in multisets $K$, which describe the lifetime of tokens. This increase is obviously the same for the entire net. Related to this is also the problem of removing elements from multiset $K$ that, due to the elapsed time, no longer satisfy the condition of belonging to $K$. This is equivalent to removing a token from a given place. The same amount of time is also added for transition timers $u_t$ and $w_t$ described in Definition \ref{def:xTPNtstate}.
	\item Creating new tokens in place $p_j$, that is, adding new elements to the corresponding multiset $K_{p_j}$. This corresponds to the end of a production phase of some transitions.
	\item Checking whether multiset $K_{p_j}$ contains some subset with certain properties to determine whether a transition is active or inactive.
	\item Removal of a subset of tokens from a place, that is, some elements of $K_{p_j}$, as a result of starting the production phase of transition.
\end{itemize}

Definition \ref{def:xTPNtimeLapse} shows how the values of all elements of the multisets $K$ are modified by time.

\begin{mydef}\label{def:xTPNtimeLapse} The elapsing of time in multisets $K$ and $M$ \\
Let $\prescript{a_j}{0}{K}_{p_j}$ be a non-empty multiset of real numbers $\mathbb{R}_{0}^{+}$ specifying the current lifetime of each token at place $p_j$, $a_j = \gamma^{U}_{p_j}$. The elapsed time $\tau$ modifies all elements of $\prescript{a_j}{0}{K}_{p_j}$, thus creating a multiset $\prescript{a_j}{0}{K'}_{p_j}$ which is denoted as $\prescript{a_j}{0}{K}_{p_j} \xrightarrow{\tau} \prescript{a_j}{0}{K'}_{p_j}$. Multiset $\prescript{a_j}{0}{K}_{p_j}^{'}$ is defined as follows: $\prescript{a_j}{0}{K'}_{p_j} = \{\kappa'_x = \kappa_x + \tau\: \vert \: \kappa_x \in \prescript{a_j}{0}{K}_{p_j} \land \kappa'_x \leq a_j\}$.\\
The elapsed time for multiset $M = \{\prescript{a_j}{0}{K}_{p_1}, \prescript{a_j}{0}{K}_{p_2}, ..., \prescript{a_j}{0}{K}_{p_n}\}$ is denoted as $M \xrightarrow{\tau} M'$, where $M' = \{ \prescript{a_j}{0}{K'}_{p_1}, \prescript{a_j}{0}{K'}_{p_2}, ..., \prescript{a_j}{0}{K'}_{p_n} \}$
\end{mydef}

New elements can be added and removed from multiset type $K$. Definition \ref{def:xTPNaddTokens} specifies how to add new elements to multiset $K_j$.

\begin{mydef}\label{def:xTPNaddTokens} Adding elements to multisets $K$ \\
Let $\prescript{}{}{K}_{p_j}$ be multiset assigned to place $p_j$, and let $v \in \mathbb{N}$ be some natural number. Adding $v$ new elements to the multiset $\prescript{}{}{K}_{p_j}$ creates new multiset $\prescript{}{}{K'}_{p_j}$ which is denoted as $\prescript{}{}{K'}_{p_j} = \prescript{}{}{K'}_{p_j} + v = \prescript{}{}{K'}_{p_j} \cup \prescript{}{}{K}^{v}$, where $\prescript{}{}{K}^{v} = \{\kappa_i = 0 : i=1,...,v\}$.
\end{mydef}

It should be remembered that $K$ are multisets; therefore, each zero in the auxiliary multiset $K^v$ defines a new distinct token created at some point in time. For example, let $\prescript{}{}{K}_{p_j} = \{0, 0, 1, 2, 3, 3\}$ and let $v = 3$. Then the operation $\prescript{}{}{K}_{p_j} + 3$ creates a new multiset $\prescript{}{}{K'}_{p_j} = \{0, 0, 0, 0, 0, 1, 2, 3, 3\}$. It should be noted that the parameters $a_j$ and $b_j$, which are usually placed on the left side of the multisets $K$ have not been given, because their values are irrelevant in this operation.

The next operation performed on the multiset $K_{p_j}$ determines how the elements of the set are removed, which represents taking tokens from each place $p \in {^{\bullet}t}$, when some active transition $t$ begins its production phase.

\begin{mydef}\label{def:xTPNremoveTokens} Removing elements from multiset $K$ \\
Let $\prescript{a_j}{0}{K}_{p_j}$ be a multiset assigned to place $p_j$, and let $v \in \mathbb{N}$. Let there exist activating subset of tokens $\prescript{a_j}{b_j}{K}_{p_j}^{act}$. 
%such that $|\prescript{a_j}{b_j}{K}_{p_j}^{act}| \geq v$ and let $a_j = \gamma^{U}_{p_j}$ and $b_j = \gamma^{L}_{p_j}$. 
Removing $v$ elements from multiset $\prescript{a_j}{0}{K}_{p_j}$, which is denoted as $\prescript{a_j}{0}{K}_{p_j} - v$, creates new multiset $\prescript{a_j}{0}{K'}_{p_j} = \prescript{a_j}{0}{K}_{p_j} \setminus K^{-}_{p_j}$ for that place, where $K^{-}_{p_j}$ is a subset representing tokens for removal in such a way that $K^{-}_{p_j} = \{\kappa_1, \cdots, \kappa_v \in \prescript{}{b_j}{K}_{p_j}^{act} \text{ and } \sum\limits_{x=1}^{v} \kappa_x \text{ is maximal }\}$. 
\end{mydef}

%max(\sum\limits_{x=1}^{v} \kappa_x)\

Multiset $K^{-}_{p_j}$ specifies the elements to be removed from multiset $\prescript{a_j}{0}{K}_{p_j}$, when a specific transition is activated. It should be noted that it is always a subset of activation set $\prescript{}{b_j}{K}_{p_j}^{act}$ (which itself is, in turn, a subset of $\prescript{a_j}{0}{K}_{p_j}$). It is always true that $\vert K^{-}_{p_j} \vert \leq \vert \prescript{a_j}{b_j}{K}_{p_j}^{act} \vert \leq \vert \prescript{a_j}{0}{K}_{p_j} \vert$. The value $v$ is actually the weight of the arc that connects $p_j$ with transition $t$.

It should be noted that there is an arbitrarily used feature of $K^{-}_{p_j}$, that is, according to Definition \ref{def:xTPNremoveTokens}, this multiset should consist only of highest numbers of multiset $K_{p_j}^{act}$. In other words, the ''oldest'' tokens will be removed from multiset $\prescript{a_j}{0}{K}_{p_j}$. Note that this assumption is only one of three possible ones. All cases for the multiset $K^{-}_{p_j}$ are listed below:

\begin{enumerate}
\item $K^{-}_{p_j} = \{\kappa_1, \cdots, \kappa_v \in \prescript{}{b_j}{K}_{p_j}^{act} \text{ and } \sum\limits_{x=1}^{v} \kappa_x \text{ is maximal }\}$ - exactly as stated in Definition \ref{def:xTPNremoveTokens}.
\item $K^{-}_{p_j} = \{\kappa_1, \cdots, \kappa_v \in \prescript{}{b_j}{K}_{p_j}^{act} \text{ and } \sum\limits_{x=1}^{v} \kappa_x \text{ is minimal }\}$ - the multiset $K^{-}_{p_j}$ contains the elements with the smallest values, corresponding to the ''youngest'' tokens at place $p_j$.
\item $K^{-}_{p_j} = \{\kappa_1, \cdots, \kappa_v \in \prescript{}{b_j}{K}_{p_j}^{act} \}$ - no imposed condition, which in practice means that any element of the set $K_j^{act}$ can be assigned to $K^{-}_{p_j}$. This, of course, applies only to the elements for which it is true that their value is equal or greater than $\gamma_{p_j}^{L}$.
\end{enumerate}

Leaving aside the third case, a practical interpretation of the difference between the first and second cases should be mentioned. It can be shown that the second case may result in a smaller number of active transitions being started, especially when there is a conflict between transitions sharing the same pre-places. This is because using only the youngest tokens will mean that the oldest tokens can, after some time, reach and exceed the lifetime limit of $\gamma^{U}_{p_j}$ and will be removed from multisets $K$ without activating anything. In contrast, there is a higher chance of starting more transitions at once in a certain time interval, when using Definition \ref{def:xTPNremoveTokens} that describes set $K^{-}_{p_j}$ as the one having the highest numbers, representing the oldest tokens at that place.

The following two operations define the rules for adding two multisets $M$ (Definition \ref{def:xTPNsetMaddition}) and subtracting two multisets $M$ (Definition \ref{def:xTPNsetMsubtraction}). These are required to define operations for adding and removing, respectively, a number of tokens in the net. The next two definitions (Definition \ref{def:xTPN_K_plus_set} and \ref{def:xTPN_K_minus_set}) that follow will introduce special types of multisets $M$ representing consumed and produced tokens.

\begin{mydef}\label{def:xTPNsetMaddition} M-addition of multisets $M$ \\
Let two multisets $M_1 = \{K_1^1, K_2^1, \cdots, K_n^1\}$ and $M_2 = \{K_1^2, K_2^2, \cdots, K_n^2\}$ be given. The operation of M-addition of two multisets creates a multiset $M' = M_1 \:\langle\: M_2$ such that $M' = \{K'_1, K'_2, \cdots, K'_n\}$, where for each $j=1,2,...n$, multiset $K'_j = K_j^1 \cup K_j^2$.
\end{mydef}

\begin{mydef}\label{def:xTPNsetMsubtraction} M-subtraction of multisets $M$ \\
Let two multisets $M_1 = \{K_1^1, K_2^1, \cdots, K_n^1\}$ and $M_2 = \{K_1^2, K_2^2, \cdots, K_n^2\}$ be given. The operation of M-subtracting multiset $M_2$ from multiset $M_1$ creates multiset $M' = M_1 \:\rangle\: M_2$ such that $M' = \{K'_1, K'_2, \cdots, K'_n\}$, where for each $j=1,2,...n$, $K'_j = K_j^1 \setminus K_j^2$.
\end{mydef}

Two elementary operations in Petri nets concern taking a certain number of tokens from places belonging to $^{\bullet}t$, and creating a certain number of tokens in places belonging to $t^{\bullet}$. In the proposed net, two definitions of specific multisets, denoted $M^{t \oplus}$ and $M^{t \ominus}$ will be given. These two multisets will have specific $K$ multisets as elements, which in turn contain numbers denoting, respectively, all the new tokens created by a transition that has completed its production phase (Definition \ref{def:xTPN_K_plus_set}) and the tokens that a transition takes from its input places when it starts its production phase (Definition \ref{def:xTPN_K_minus_set}).

\begin{mydef}\label{def:xTPN_K_plus_set} Multiset $M^{t \oplus}$ \\
Let $t \in T$ be a transition in $\mathcal{Z}$ and $W(t, p_j)$ be the weight of the arc going from $t$ to place $p_j$. Multiset $M^{t \oplus} = \{K^{t \oplus}_1, \dots, K^{t \oplus}_{|P|}\}$ consists of multisets $K^{t \oplus}_j$ such that:
\[
 \forall p \in P : K^{t \oplus}_{p_j} = \begin{cases}
\{0\: |\: 0 \text{ appears } W(t, p_j) \text{ times} \} \;\; \textit{if} \:\: p_j \in t^{\bullet} \\
\emptyset \quad\quad\quad\quad\quad\quad\quad\quad\quad\quad\quad\quad \textit{if} \:\: p_j \notin t^{\bullet} \\
\end{cases}
\]
\end{mydef}

The multiset $M^{t \ominus}$ is introduced by Definition \ref{def:xTPN_K_minus_set}.

\begin{mydef}\label{def:xTPN_K_minus_set} Multiset $M^{t \ominus}$ \\
Let $t \in T$ be a transition in $\mathcal{Z}$ and for each place $p_j \in ^{\bullet} t$ there exist activating subset of tokens $\prescript{a_j}{b_j}{K}_{p_j-t}^{act}$. Let $W(p_j, t)$ be the weight of the arc directed from $p_j$ to $t$. Multiset $M^{t \ominus} = \{K^{t \ominus}_1, \dots, K^{t \ominus}_{|P|}\}$ consists of multisets $K^{t \ominus}_{p_j}$ such that: 
\[
\forall p \in P : K^{t \ominus}_{p_j} = \begin{cases}
%\{\kappa_1, \cdots, \kappa_{W(p_j, t)} \in \prescript{a_j}{b_j}{K}_{p_j}^{act} \text{ and} \sum\limits_{x=1}^{W(p_j, t)} \kappa_x \text{ is maximal}\} \quad \textit{if} \:\: p_j \in {^{\bullet}}t \\
\{\text{contains exactly } W(p_j, t) \text{ maximal numbers from } \prescript{a_j}{b_j}{K}_{p_j-t}^{act}\} \textit{ if } p_j \in {^{\bullet}}t \\
\emptyset \quad\quad\quad\quad\quad\quad\quad\quad\quad\quad\quad\quad\quad\quad\quad\quad\quad\quad\quad\quad\quad\quad\quad\quad \textit{ if } p_j \notin {^{\bullet}}t .\\
\end{cases}
\]

\end{mydef}

In Definition \ref{def:xTPN_K_plus_set} multiset $M^{t \ominus}$ is described. Each multiset $K^{t \ominus}_{p_j} \in M^{t \ominus}$ (if not empty) describes new tokens (given by zero) that will be added to the net if some transition $t$ completes its production phase (obviously, $|K^{t \ominus}_{p_j}| = W(t, p_j)$. For places $p_j \notin {^{\bullet}}t$, multiset $K^{t \ominus}_{p_j} = \emptyset$.

Similarly, Definition \ref{def:xTPN_K_minus_set} specifies multisets of tokens to be removed from the net, when some transition $t$ starts its production phase. These tokens are described by the corresponding numbers included in multisets $K^{t \ominus}_{p_j}$, for all places $p_j \in {^{\bullet}{t}}$. For the remaining places, these are empty sets. It should be noted that, as already specified in Definition \ref{def:xTPNremoveTokens} these are the elements that represent the oldest tokens. Therefore, $W(p_j, t)$ highest numbers are assigned to $K^{t \ominus}_{p_j}$ from activating subset $\prescript{a_j}{b_j}{K}_{p_j-t}^{act}$.

\section{State changes}

In the proposed net, each change in state depends on the passage of time. In this section a formal definition of state changes will be given. Before formal definitions for state changes are given, we will here summarize the basic semantics of the proposed net.

\begin{itemize}
\item \textbf{Activation phase.} Transition $t_i$ becomes active (denoted as $t^{\#} \rightarrow t^{act}$) if in all of its input places there exists an activating subset of tokens (Defition \ref{def:xTPNtransitionActive}). Upon activation, counter $u_i$ for transition $t_i$ is initialized. It will count from zero towards some value within the range [$\alpha^{L}_{i}, \alpha^{U}_{i}$]. The value of $u_i$ increases with time $\tau$. When the transition state changes from activation to production, such a change is denoted as $t^{act} \rightarrow t^{\#} \lor t^{prod}$.

\item \textbf{Production phase.} Transition $t_i$ consumes tokens from its input places (based on the arc weights). Counter $w_i$ for transition $t_i$ is initialized and will count from zero towards some value within the range [$\beta^{L}_{i}, \beta^{U}_{i}$]. The value of $w_i$ increases with time $\tau$. When the transition finally produces tokens at all places from set ${t_i}^{\bullet}$, it becomes active or inactive depending on the existance or not of activation subsets (such a change is respectively denoted as $t^{prod} \rightarrow t^{\#}$ or $t^{prod} \rightarrow t^{act}$. A special case can be observed when the activation interval for a transition is zero. In such a case, if there exists activating subsets of tokens, such a transition starts its production phase again, by immediately taking new tokens. Such specific transition will be described later.

\item \textbf{Transition states.} Function $h$ assigns t-state of each transition $t_i$ with the ordered pair $h(t_i) = (u_i, w_i)$ (Definition \ref{def:xTPNtstate}).

\item \textbf{Token dynamics}. Each token in a place has distinct lifetime counter $\kappa_x$ that starts at zero when a token is created in a place. Lifetime $\kappa_x$ of each token increases with the flow of time $\tau$. A token matures when its lifetime is equal to or exceeds maturity value $\gamma^{L}_j$ of place $p_j$. A token is removed from its place when its lifetime exceeds maximum value $\gamma^{L}_j$ of that place.

Regarding tokens, as explained in the description of Definition \ref{def:xTPN_K_minus_set}, it is a matter of exactly which tokens will be taken from the input places at the start of the production phase. At the beginning of the production phase, transition $t$ takes the appropriate number of tokens from its input places. Three ways telling which tokens can be taken have been described as a comment to Definition \ref{def:xTPNremoveTokens}: the youngest, oldest, or random ones. It is assumed for the proposed net that the oldest tokens are consumed when production phase starts for any transition. 
%Of course, each token consumed must have a lifetime equal to at least $\gamma^{L}_{p}$. 
%Taking only the oldest tokens seems to be the optimal approach to maximize the overall transition firing in the net. If the youngest or random tokens are taken when production starts, there may be states in which the oldest tokens in some places exceed their maximum allowed lifetime $\gamma^{U}_{p}$, before they are used by any of the output transitions. This can result in the disappearance of activating subsets for some transitions and their consequent deactivation. In summary, it can be assumed that the way in which tokens are consumed when a transition begins production depends on the needs of a particular model. However, taking the oldest ones seems to be a better default strategy. 
\end{itemize}

When the p-state of the net is described at a given time (by multiset $M$), and if there exists transition $t$ that is active and can start producing tokens, one can say that:
\begin{itemize}
	\item Transition $t$ will take a number of tokens from the places resulting in transforming multiset set $M$ into $M'$ ($M \rightarrow M'$), where $M' = M \:\rangle\: M^{t \ominus}$ (Definition \ref{def:xTPNsetMsubtraction}),
 \item Then, after some time, $t$ will produce new tokens (represented in multiset $M^{t \oplus}$), transforming multiset $M'$ into $M''$, ($M' \rightarrow M''$) where $M'' = M' \:\langle\: M^{t \oplus}$ (Definition \ref{def:xTPNsetMaddition}).
\end{itemize}

Some time must pass between the state described by the multiset $M'$ and the state described by $M''$ (this is the so-called token production time). However, from the moment the production begins, there is no possibility to interrupt this process. A transition that is already in the production phase will always create new tokens after its production time.

The activation state of the transition can be stopped but only if, at any moment of time, for some of its input places, a required activation subset of tokens vanishes. Unlike other popular time Petri nets, changing the state of other transitions being in conflict (i.e., sharing common input place) does not reset the activation timer (in xTPN represented by $u_t$). For example, if $t_1$ and $t_2$ share the same place $p_1$, when $t_1$ starts or ends its production phase and there is still an activating subset of tokens for $t_2$ in $p_1$, timer $u_{t_2}$ does not reset. It still counts time towards the moment when the production phase start. 
%However, there is a possibility, where the reset of the timer for $t_1$ seems appropriate, in a situation where the conflicted transition changes the number of tokens in the shared input place ($t_1$ in an example). When determining $\tau^{\alpha}$ and $\tau^{\beta}$ for both $t_1$ and $t_2$, a \emph{mass action law} can be considered. In simple words, using this law in Petri net simulation, assume that the higher the number of tokens there are in a given transition input places, the higher the firing rate of such a transition should be. Therefore, in this scenario, the initial activation or production can be faster if there is an abundance of tokens in $p_1$. One can say that in a hypothetical \emph{multiple server semantics} such a transition could be triggered multiple times in a given moment. Proposed xTPN follows \emph{single-server semantics} (i.e., transition fires once). However, if the mass action law is considered, the change in the number of tokens in the input places can be considered as a sufficient reason to reset its activation timer.

\subsection{Definitions for xTPN state change}

The state of an xTPN consists of two sub-states: $p$-state and $t$-state (Definitions \ref{def:xTPNpstate} and \ref{def:xTPNtstate}), which together form full state $z = (m, h)$. The $p$-state is a function $m$ that assigns to each place multiset $K$, whose elements describe the lifetime of each token in a given place. All these multisets together form multiset $M$. The second substate, $t$-state, is described by function $h$ that assigns to each transition a pair consisting of real numbers and/or a special symbol $\#$. 

Changing the state of $\mathcal{Z}$ is possible in three ways. The first is the passage of time. As an immediate effect, elements of multisets $K$ are modified in such a way that each of them is incremented by the same value  $\tau$ representing the time that has passed. Therefore, some tokens may reach an adult form (defined independently for each place by value $\gamma^{L}_p$), which as a result may allow an activation of some transitions. Some tokens may also exceed value of $\gamma^{U}_p$. As a result they will be immediately removed from a given place, that is, from the corresponding multiset $K_p$. This in turn may result in a deactivation of some transitions. The passage of time also directly affects all transitions, i.e., their timers, $u$ or $w$, are increased by the same value $\tau$. These changes are described by Definition \ref{def:xTPNstateChangeTimeElapse}.

\begin{mydef}\label{def:xTPNstateChangeTimeElapse} Change the net state of $\mathcal{Z}$ by elapsing time.\\
Let net $\mathcal{Z}$ be in state $z = (m, h)$ and positive number $\tau$ be given. Multiset $M$ contains multisets $K$ describing tokens in places. After a time lapse of $\tau$, the net changes its state $z$ to $z'$, which we write as $z \xrightarrow{\tau} z'$, and state $z' = (m', h')$ is described as follows:
\begin{enumerate}
\item $M' = M \xrightarrow{\tau} M'$
\item $m' : T \rightarrow M'$, where  
\item $\forall t : h'(t) = \begin{cases}
 (\#, \#) \quad\quad\quad\; \textit{if } \quad M^t \not\subset M' \land w_t = \#\\
 (\#, w_t + \tau) \quad\; \textit{if } \quad w_t \neq \#\\
 (u_t + \tau, \#) \quad\;\; \textit{if } \quad M^t \subset M' \land M^t \subset M \land w_t = \#\\
 (0, \#) \quad\quad\quad\;\; \textit{if } \quad M^t \subset M' \land M^t \not\subset M \land w_t = \# 
\end{cases}$
\end{enumerate}
\end{mydef}

The state of the net is defined by pair $z := (m, h)$. p-state is given by the function $m$, which for each place assigns elements of multiset $M$, such that $m(p) = \prescript{\gamma^{U}_{p}}{0}{K}_{p}$ (Definition \ref{def:xTPNpstate}). The $t$-state is given by function $h$, which for each transition assigns a pair of two values (timers) such that $h(t) = (u_t, w_t)$ (Definition \ref{def:xTPNtstate}).

Definition \ref{def:xTPNstateChangeTimeElapse} precisely tells how the state $z$ changes, in three consecutive steps. The first two steps specify a change in $p$-state (that is, for net places). At first, multiset $M$ is transformed into $M'$ (Definition \ref{def:xTPNtimeLapse}). Then, multisets $\prescript{\gamma^{U}_{p}}{0}{K'}_{p} \in M'$ are assigned to each place $p$. Each such multiset is created by increasing its elements by value of $\tau$.

Changes in token lifetimes can affect the activation state of a transition. The third step describes four possibilities to update the $t$-state of a given transition $t \in T$. Multiset $M^t$ (described in Definition \ref{def:xTPNtransitionActive}) is determined for each transition $t$. Each of these four cases will now be explained.

The first case is valid for transitions that are not producing anything at the moment ($w_t = \#$) and are not active in the new $p$-state, that is $M^t \not\subset M'$ (cf. Definitions \ref{def:xTPNkinclusion} and \ref{def:xTPNtransitionActive})

%If a given transition in a $t$-state described by multiset $M$ has been active, then $M^t$ contains elements representing tokens from input places of $t$ that maintain the activation state. However, if any input place $p$ of $^{\bullet}{t}$ lacks a sufficient number of tokens with lifetimes that have values greater than or equal to $\gamma^{U}_{p}$, the multiset $K_p \in M^t$ will contain only the special element $\#$, resulting in $M^t \nleq M'$. This means that $t$ is not active in the new $p$ state described by the multiset $M'$, therefore $h'(t) = (\#, \#)$. For $t$ and its $t$-state function $h(t) = (u_t, w_t)$ it is necessary that $w_t = \#$ meaning $t$ has not produced anything in the $p$-state $m$.

The second case is valid for transitions that produced tokens in the previous $t$-state, that is, if $w_t \neq \#$, when $w_t \in \mathbb{R}^{+}_{0}$. For these transitions, their production timer $w_t$ increases over time $\tau$.

The third case applies to all active transitions $t$ in the new $p$-state ($M^t \subset M'$) that were already active in the previous $t$-state (meaning $M^t \subset M$) and in that previous state did not produce anything ($w_t = \#$). Transitions in this group update their activation timer $u_t$ by time $\tau$. 

Finally, it is important to note that aging tokens over time $\tau$ can lead to an activation of transitions. The fourth and final case of step 3 of Definition \ref{def:xTPNstateChangeTimeElapse} deals with this situation. If a transition is active in the new $p$-state described by multiset $M$ (i.e. $M^t \subset M'$), but was previously neither active ($M^t \not\subset M$) nor produced tokens (i.e., $w_t = \#$), then function $h'$ assigns to it pair $(0, \#)$. In simpler words, such a transition starts its activation phase as a result of time $\tau$ passing for the net.

The second way of changing the state of the net occurs when a transition starts producing tokens. It involves taking a certain subset of tokens from places. Definition \ref{def:xTPNstateChangeProdStart} given below describes how exactly the state of the net changes when a certain transition starts the token production phase (denoted as $t^{act} \rightarrow t^{prod}$).

\begin{mydef}\label{def:xTPNstateChangeProdStart} Change a state of $\mathcal{Z}$ by starting the production phase \\
Let a net $\mathcal{Z}$ be in state $z = (m, h)$ in which a specific transition $\hat{t}$ starts its production phase, which is denoted as $\hat{t}^{act} \rightarrow \hat{t}^{prod}$. As a result, the state of $\mathcal{Z}$ changes to $z' = (m', h')$ (denoted $z \xrightarrow{\hat{t}^{act} \rightarrow \hat{t}^{prod}} z'$). New state $z'$ is described as follows:
\begin{enumerate}
\item $M' = M \:\rangle\: M^{\hat{t} \ominus}$
\item $m' : P \rightarrow M'$ 
\item $\forall t : h'(t) := \begin{cases}
 (\#, \#) \quad\: \textit{if} \:\: M^t \not\subset M' \land w_t = \#\\
 (0, \#) \quad\;\; \textit{if} \:\: M^t \subset M' \land M^t \subset M \land w_t = \# \\
 h(t) \quad\quad\: \textit{otherwise.}
\end{cases}$
\end{enumerate}
\end{mydef}

As before, there are three steps in which the state changes. The first and second steps describe the change in the $p$-state. First a new multiset $M'$ is created, by removing tokens from input places of $\hat{t}$ which starts its production phase. These tokens are described by a multiset $M^{\hat{t} \ominus}$ (Definition \ref{def:xTPN_K_minus_set}). Then, in step two, new multiset $M'$ starts describing new $p$-state of the net.

The third step describes the $t$-state update. It has three possible cases how function $h'(t)$ can describe new $t$-state of each transition.

The first case concerns transitions that were not producing anything in the previous $p$-state and are now inactive in the new $p$-state defined by $M'$. 

The second case concerns transitions that may become active in a new state, when $\hat{t}$ starts production phase. Activation of some other transitions because other one has taken tokens is possible if inhibitor arcs exist in $\mathcal{Z}$. Such an arc will be explained later, but if it exists, taking tokens from an input place of $\hat{t}$ may result in the activation of some other transitions. 

The third and last case states that every other transition timer remains unchanged. This applies to transitions that were already active in the previous $p$-state described by $M$ or when they are still producing (indicated by $w_t \neq \#$).

Definition \ref{def:xTPNstateChangeProdEnd} describes the state change as a result of finishing the production phase by any transition. 
%A transition that initiates that change is the one for which $w_{t} = 0$, i.e., its production timer stopped counting towards zero from the initial value $\tau^{\beta}_{t}$ (where $\alpha^{L}_{t} \leq \tau^{\beta}_{t} \leq \alpha^{U}_{t}$).

\begin{mydef}\label{def:xTPNstateChangeProdEnd} Change the state of $\mathcal{Z}$ by ending the production phase \\
Let a net $\mathcal{Z}$ be in state $z = (m, h)$, in which transition $\hat{t}$ finishes its production phase, which is denoted as $\hat{t}^{prod} \rightarrow \hat{t}^{\#}$ or $\hat{t}^{prod} \rightarrow \hat{t}^{act}$. As a result, the state of $\mathcal{Z}$ changes to $z' = (m', h')$ 
%(denoted either as $z \xrightarrow{\hat{t}^{prod} \rightarrow \hat{t}^{\#}} z'$ or $z \xrightarrow{\hat{t}^{prod} \rightarrow \hat{t}^{\#}} z'$). 
The new state $z'$ is described as follows:
\begin{enumerate}
\item $M' = M \:\langle\: M^{\hat{t} \oplus}$
\item $m' : T \rightarrow M'$ 
\item $\forall t : h'(t) := \begin{cases}
 (\#, \#) \quad\: \textit{if} \:\: M^t \not\subset M' \land w_t = \#\\
 (0, \#) \quad\;\; \textit{if} \:\: M^t \subset M' \land M^t \not\subset M \land w_t = \#  \\
 h(t) \quad\quad\: \textit{otherwise.} 
\end{cases}$
\end{enumerate}
\end{mydef}

As one can see, this definition is almost identical to Definition \ref{def:xTPNstateChangeProdStart}. The most important difference is in its first of three steps, describing how the net state $z$ changes. The new $p$-state is introduced and described by the multiset $M'$. This multiset is created by adding to elements of $M$ elements of multiset $M^{\hat{t} \oplus}$ (Definition of \ref{def:xTPN_K_plus_set}). Multiset $M^{\hat{t} \oplus}$ contains data on all new tokens that will be created, when $\hat{t}$ finishes its production state. The rest of Definition \ref{def:xTPNstateChangeProdEnd} is the same as in Definition \ref{def:xTPNstateChangeProdStart}, where the start of the production phase has been described. It should be noted that after finishing the production, transition $\hat{t}$ may become active or inactive. It depends on its activation subsets in $p$-state described by $M'$. 

\subsection{Remarks on state changes in net simulation}

In the proposed net $\mathcal{Z}$ state $z$ changes with each passing time unit $\tau$, which we denote as $z \xrightarrow{\tau} z'$. In theory, there can be an infinite number of states between $z$ and $z'$, since $\tau$ can be any real number. Activation and production times of a transition ($\tau^{\alpha}_t$ and $\tau^{\beta}_t$ respectively) are random with respect to alpha and beta ranges. Let us however assume, that times $\tau^{\alpha}_t$ and $\tau^{\beta}_t$ ($\alpha^{L}_t \leq \tau^{\alpha}_{t} \leq \alpha^{U}_t$ and $\beta^{L}_t \leq \tau^{\beta}_{t} \leq \beta^{U}_t$) are known. Such an assumption can be true for example in a hypothetical simulator of the proposed net, which must establish these values at some given moment in order to change net state. Four so called \emph{relevant states} that can be described as follows: 
\begin{enumerate}
 \item A state when any token achieves lifetime equal to $\gamma^{L}_p$ in any place $p \in P$.
 \item A state when a token in any place $p \in P$ exceeds by any value the maximal allowed lifetime for that place, that is, the token lifetime becomes greater than $\gamma^{U}_p$.
 \item A state when for any transition $t \in T$ its activation timer $u_t$ reaches the value $\tau^{\alpha}_t$, that is, when the active transition starts the production phase.
\item A state when for any transition $t \in T$ its production timer $w_t$ reaches value $\tau^{\beta}_t$, that is, when the producing transition finally creates tokens.
\end{enumerate}

We called these four states the relevant ones, because any other net state associated with changing time by $\tau$ value only do just that, i.e., it will update time values in net elements, but without changing, e.g., the number of tokens anywhere or causing any transition to switch between its possible $t$-states.

Assuming that times $\tau^{\alpha}_t$ and $\tau^{\beta}_t$ are computed and known, when the $t$-state for a transition changes, it is always possible to calculate next time $\tau$ to the one of the four relevant states enumerated previously. Four possible values of these time can be denoted as $\tau^{mat}$, $\tau^{exp}$, $\tau^{act}$ and $\tau^{prod}$. They are time for (relevant) states when token matures, time for token expiration, time when transition activation ends, and finally time when transition production phase ends, respectively. In Table \ref{tab:relStates} conditions are given that allow identification of such states. 

\begin{table}[ht]
	\begin{center}
		\caption{Relevant states achieved after certain time and possible changes in state description.}
		 \label{tab:relStates} 
		\begin{tabular}{ |p{4cm}|p{4.5cm}|p{6.5cm}|  }
		 %\hline
		 %\multicolumn{3}{|c|}{Possible transformations} \\
		 \hline
		 Relevant state of $\mathcal{Z}$ achieved after time $\tau$ & Conditions & Possible effects on $t$-state and $p$-state. \\
		 \hline \hline
		 Maturity state for tokens after time $\tau^{mat}$  & There exists a token in place $p$, described by $\kappa \in K_p$, such that $\kappa + \tau^{mat} = \gamma^{L}_p$ & 
        Transition $t$ may become active. When inhibitor arc exists, a transition may become inactive.
   %$t^{\#} \rightarrow t^{act}$ \newline $t^{act} \rightarrow t^{\#}$ (for inhibitor arc) 
        
   \\ \hline
	Token expire after time $\tau^{exp}$  & There exists a token in place $p$, described by $\kappa \in K_p$, such that $\kappa + \tau^{exp} > \gamma^{U}_p$ & 
        Transition $t$ may become inactive. When inhibitor arc exists, a transition may become active.
   %$t^{act} \rightarrow t^{\#}$ \newline $t^{\#} \rightarrow t^{act}$ (for inhibitor arc) 
        \\ \hline
    A state when active transition starts production phase after $\tau^{act}$  & There exists transition $\hat{t}$ for which $u_{\hat{t}} - \tau^{act} = 0$ & 
   %$\hat{t}^{act} \rightarrow \hat{t}^{prod}$ \newline $t^{act} \rightarrow t^{\#}$ \newline $t^{\#} \rightarrow t^{act}$ (for inhibitor arc) \newline removal of tokens in $^{\bullet}\hat{t}$ 
        $\hat{t}$ changes state from active to producing.  Any other active transition $t$ may become inactive, and if inhibitor arc exists, an inactive transition may become active. Tokens are removed from places in $^{\bullet}\hat{t}$.
        \\ \hline
	A state when a transition ends its production phase after $\tau^{prod}$  & There exists transition $\hat{t}$ for which $w_{\hat{t}} - \tau^{prod} = 0$ & 
    %$\hat{t}^{prod} \rightarrow \hat{t}^{\#} \lor \hat{t}^{act}$ \newline $t^{\#} \rightarrow t^{act}$ \newline $t^{act} \rightarrow t^{\#}$ (for inhibitor arc) \newline creating tokens in $\hat{t}{^{\bullet}}$ 
    $\hat{t}$ changes state from producing to active or inactive.  Any other inactive transition $t$ may become active, and if inhibitor arc exists, an active transition may become inactive. Tokens are created in place belonging to $\hat{t}{^{\bullet}}$.
    \\ \hline
		\end{tabular}
	\end{center}
\end{table}

The time to the nearest relevant state can be calculated, i.e., it is a minimal one of $\tau^{mat}$, $\tau^{exp}$, $\tau^{act}$ and $\tau^{prod}$. After exactly that time, new achieved state $z'$ can be characterized as a relevant one. Any time smaller than such a value will of course change the current state of the net, but only in such a way that it will increase token lifetimes and timers of transitions. No other changes will occur in such a case.

By analyzing the elements of all multisets $K$, two time values that could lead to a new relevant state of $\mathcal{Z}$ can be calculated. The first is maturity time $\tau^{mat}$ and is the time after which a certain token will reach a lifetime equal to $\gamma^{L}_{p}$ and can then potentially be used in some activating subset. The second time value is $\tau^{exp}$ (expiration time) and it is the minimum time after which a certain token will exceed $\gamma^{U}_{p}$ value and will be removed from place $p$. This can potentially change the $t$-states of some transitions. Similarly, two time values $\tau^{act}$ and $\tau^{prod}$ can be calculated by analyzing transition timers. The first value is the time exactly after which certain transition $t$ ends its activation phase, i.e., a value for which $u_t - \tau^{act} = 0$. When that happens, tokens must be taken from such transition input places. The last of the four time values is $\tau^{prod}$ and it is a time after which a certain transition $t$ finishes its production phase, i.e.,  $w_t - \tau^{prod} = 0$. In that case, some new tokens will also be added to the transition output places.

Changes possible in a relevant state are given in the last column of Table \ref{tab:relStates}, and some of them are only possible when inhibitor arcs exist in a given net. For example, when a token reaches maturity age, it can potentially allow activation of some transition (that potential change in the new relevant state is described in the first row of the table as $t^{\#} \rightarrow t^{act}$). However, when an inhibitor arc exists, a token that reaches maturity age can potentially trigger that inhibitor arc, which in turn causes immediate deactivation of an active transition.

\section{Transformations and specific arcs}

In this section, two interesting features of the proposed net will be described. The first of them is the ability to transform net elements into more simple ones, known from the time Petri net literature. Later in this section we will describe some interesting features of read arcs and inhibitor arcs, if they are to be used in the proposed net.

\subsection{Places and transitions transformations}

One of potentially very useful features of the described Petri nets is the possibility of transforming its elements (both places and transitions) to the ones that are known in other types of time Petri net. This makes it possible to create a model of a system in which different types of transitions and places with time parameters can exist simultaneously. Even more important is the fact that different types of time data that describe such a system can be used in a single model without the need of additional adjustments. In Table \ref{tab:transform} possible transformations are given, which depend on the proper assignment of values to specific time variables $\alpha$, $\beta$ and $\gamma$.

\begin{table}[h]
	\begin{center}
		\caption{Possible transformations of extended time Petri net elements into simpler ones.}
		\label{tab:transform} 
		\begin{tabular}{ |p{2.2cm}|p{7cm}|p{3.5cm}|  }
		 %\hline
		 %\multicolumn{3}{|c|}{Possible transformations} \\
		 \hline
		 Net element & Time parameters settings & Result \\
		 \hline
		 Place   & $\gamma^{L} = 0$, $\gamma^{U} = \infty$    & Classical place \\
		 Transition   & $\alpha^{L}, \alpha^{U} \in {\mathbbmsl{Q}}_{0}^{+}$, $\beta^{L} = \beta^{U} = 0$    & TPN transition \\
		 Transition   & $\alpha^{L} = \alpha^{U} = 0$, $\beta^{L}, \beta^{U} \in {\mathbbmsl{Q}}_{}^{+}$  & ITPN transition \\
		 Transition   & $\alpha^{L} = \alpha^{U} = 0$, $\beta^{L}, \beta^{U} \in {\mathbbmsl{Q}}_{}^{+}$, $\beta^{L} = \beta^{U}$  & DPN transition \\
		 Transition   & $\alpha^{L} = \alpha^{U} = 0$, $\beta^{L} = \beta^{U} = 0$  & Classical transition \\
		 \hline
		\end{tabular}
	\end{center}
\end{table}

To transform the xTPN time place into a classical one (i.e., without time constraints for tokens), $\gamma^{L}_{p} = 0$ and $\gamma^{U}_{p} = \infty$. Tokens in such a place will never expire and they can be used to activate a transition immediately when they are created. Therefore, only their number matters, not their lifetimes.

DPN network can be obtained, when both $\alpha$ values are zero, while both $\beta$ variables are equal to positive values. Assuming that all places are classical as explained, all transitions start producing tokens immediately after they become activated, while the production time is equal to $\beta$ value, which is the exact behavior of the duration Petri net. The interval-time Petri net is obtained almost in the same way, i.e., in that case $\beta$ values are not equal and therefore define some production time interval.

An interesting situation occurs when zeros are assigned to the variables $\alpha$ and $\beta$. Theoretically, a classical transition is obtained. However, there is a problem in interpretation, whether such a transition should behave like in a classical Petri net, or is it rather an immediate transition, firing as soon as possible (since both time are zero).

Following firmly the time rules of $\mathcal{Z}$, this should be an immediate time transition, that activates instantly if there are enough tokens in its input places (because $\tau^{\alpha} = 0$) and also produces tokens immediately ($\tau^{\beta} = 0$). Therefore, such a transition fires immediately as many times as all tokens are consumed from its input places. 

This, of course, may be the expected behavior, but in a other simulation of $\mathcal{Z}$ one can distinguish between an immediate time transition and an ordinary classical Petri net transition. For the latter, additional assumptions can be made that for example it will fire with $50\%$ probability if it is active and only when state of the net changes from one relevant state to the other.

Another important remark should be made when a transition with zero activation and production time is considered as immediate time transition (as opposed to the classical Petri net one). In such a case, it should not be the input transition of the net (i.e., the one with no input places), because it is then forced to fire immediately and infinitely, which is probably not the desired behavior. When it is not an input transition, then it will eventually run out of tokens in its input places. 
%Assuming that the net has time-places with $\gamma$ values, even immediate transition must wait some time, before the necessary number of mature tokens becomes available.

\subsection{Read arcs and inhibitor arcs}

A read arc in a Petri net leads in both directions: to/from a place and transition. This means that the same number of tokens that have been taken from a place are returned to it by the same transition. Since in a classical Petri net the firing of an active transition is an instantaneous event, one can also say that the number of tokens in the place connected by the read arc does not change. The final state (i.e., tokens distribution) in both cases is always the same. 

A read arc can also be graphically represented by two arcs with the same weight between the same place and transition but pointed in opposite directions. Such a specific arc can also be called a \emph{double arc} and its two unidirectional arcs can then in theory have different weights. In this section we will assume the weights are exactly the same (i.e., in simple words, we will deal with a read arc) but in the further example a graphical form of a double arc will be used (yet with both arcs having exactly the same weight).

When using read arc in an extended time Petri net, a possibility appears of different behaviors of such an arc. The first one described below is assumed to be a default one for the proposed net. However, there is a least a theoretical possibility to consider some interesting different behaviors which we want briefly present in this section. Their existence proves that read arc in the proposed net can be a different construct that in the classical Petri nets.

\begin{figure}[ht!]
\centering
    \includegraphics[width=0.7\textwidth]{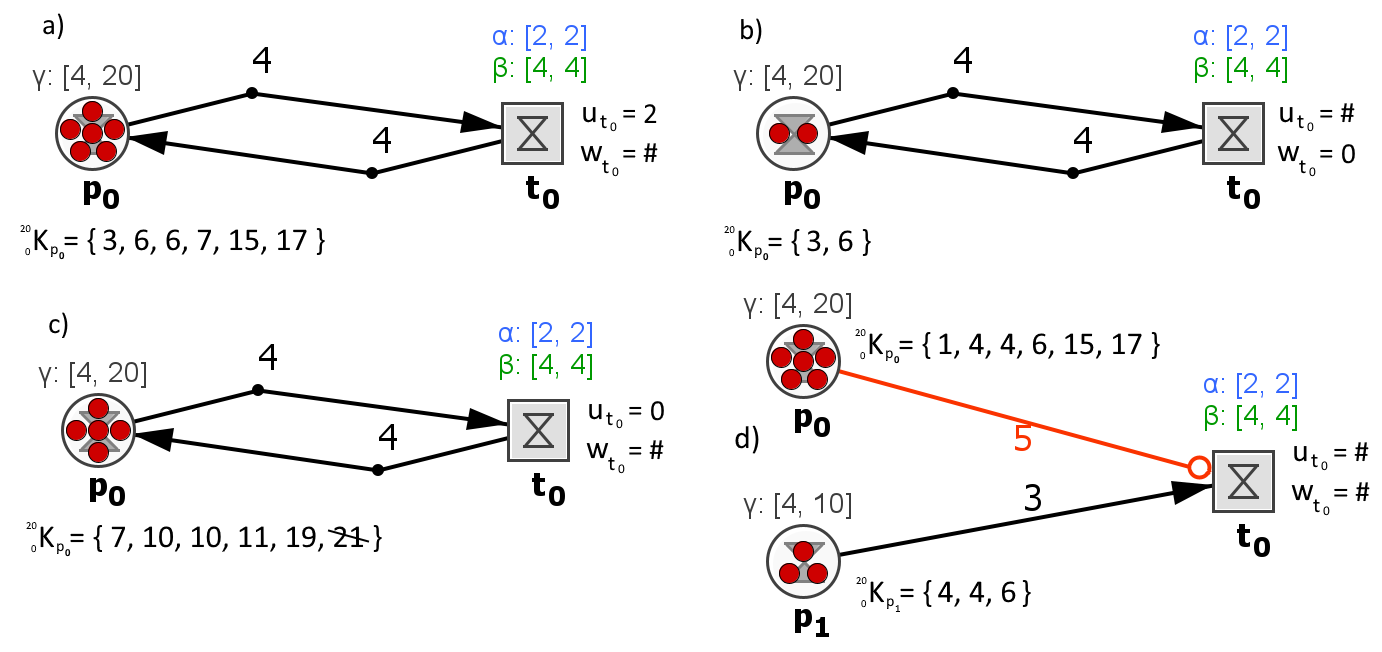}
    \caption{Part a) b) and c) represents three stages of a scenario in which tokens are returned by a read arc, but with their lifetimes increased. Part d) represents a simple example of the behavior of inhibitor arc painted in red.}
		\label{fig:4}
\end{figure}

\begin{enumerate}
\item When a production phase starts for a given transition $t$, tokens from places connected by read arcs \emph{are not token}. Obviously, they \emph{will not} be produced in any place connected by a read arc with its input transition $t$ when its production phase ends. In this way, tokens are only used to maintain the activation phase of $t$ and do not participate in its production phase. This is a straightforward and proposed default behavior for the extended time Petri net. It is closest to the behavior of a read arc in other types of Petri nets, with one important difference that tokens in place connected by such a read arc are not used or considered for anything by transition $t$ during its production phase.

\item There is however a possibility for different behaviors of an read arcs in xTPN. Their existence do not contradict the presented time rules of the net, however, ''read arc'' of such type start behave much differently than the ones in a classical nets. When the production phase starts, tokens are taken from places connected by read arcs. Therefore, such tokens cannot be used to maintain an activation of some other transitions if they share such common input places (since such tokens are for time being taken from the respective places). Conversely to the previously described default behavior in the first point, such tokens participate in the production. These tokens will obviously return in the same number after production time $\tau^{\beta}_t$ has passed. However, an interesting problem appears here, that is, in a question with not such obvious answers, what lifetime values these tokens should have. Below we present two theoretical possibilities.
\begin{enumerate}[label=(\roman*)]
\item Tokens are created with a lifetime equals to zero in a place connected by read arc, from which they were taken previously. This follows strictly from the definitions and rules presented in this paper, that each newly created token starts with lifetime set as zero.
\item Tokens are returned having their lifetimes increased by $\tau^{\beta}_{t}$. The rationale for such a scenario comes from the fact, that tokens are distinct distinguishable objects with their own lifetimes. Taking such object to do its function during production which takes $\tau^{\beta}_{t}$ time units, should also have some influence on its (token) lifetime. The previous point assume 'renewal' of lifetime of returned tokens, by using connections with read arc. This scenario however, conserves token lifetime in a given place, and it can have some additional and surprising consequences as presented in an example in Figure \ref{fig:4}. 
\end{enumerate}
\end{enumerate}

An example net in part a) of Figure \ref{fig:4} is in a state, where transition $t_0$ is already activated and ready to start token production, because its activation timer $u_{t_0} = 2$. When the transition became active exactly 2 time units before the state given in part a), its activation timer was initially set to 0. Therefore, in part b) of Figure \ref{fig:4} transition $t_0$ has taken four oldest tokens (represented by elements 6, 7, 15 and 17 of $ \prescript{20}{0}{K}_{p_0}$), its activation timer has been set to $u_{t_0} = \#$ and its production timer $w_{t_0}$ has been set to 0. Finally, as shown in part c) of Figure \ref{fig:4}, four time units have passed and the production phase of $t_0$ has ended. Four tokens must be returned to place $p_0$ with their lifetimes increased by $\tau^{\beta} = 4$ time units. The oldest of the returned tokens has now a lifetime equal to 21. This value cannot be stored in $ \prescript{20}{0}{K}_{p_0}$, therefore, effectively only 3 tokens have been returned, when four have been taken. As one can see, assuming that token lifetime is conserved by using read arc, this results in a situation, where after time $\tau^{\beta}$, the number of returned tokens in a place is lower than it has been before firing of a transition.

To avoid such a situation, the default assumed behavior for xTPN is the one when tokens are not taken using read arc. Therefore they do not participate in the production phase and can be used for other purposes. Alternatively, it could be assumed that tokens will be consumed using read arc, when a transition starts its production phase, but after that, they are returned with their lifetimes set to zero. This is also in accordance with the rules how tokens are produced in the proposed time net.

%Exactly in what way a read arc in an extended time Petri net should behave, in terms of returned tokens and their lifetimes, will most probably depend on the model. There is one more interesting feature of a read arc in $\mathcal{Z}$, if the above-mentioned behavior is assumed, i.e., if tokens are returned with lifetimes increased by the production duration of a transition that took them. When normal arcs are used, it has already been explained that in order to maximize tokens ``usefulness'' in a net, the oldest ones should be taken from a place when a production phase starts for a given transition. From this perspective, the opposite strategy can be assumed for read arcs. If the oldest tokens are taken, there is a greater probability that some of them will be returned with lifetimes already exceeding the limit value $\gamma^{U}_p$ for a given place, thus they will be removed immediately from a place as in the example in Figure \ref{fig:4}. Therefore, when using read arcs in an extended time Petri net, in order to minimize the number of such situations, the youngest tokens (i.e., the ones with lifetimes equal to or greater than $\gamma^{L}_p$) should be taken.

At the end of this section we will briefly discuss another type of non-classical arc that can exist in a Petri net, i.e., the inhibitor arc. Such an arc blocks the possibility of activating a transition towards which it is directed. In a classical Petri net, when an inhibitor arc leads from a place $p$ towards transition $t$, the latter is disabled if in place $p$ there are at least $W(p, t)$ tokens. In the discussed net, there can exist inhibitor arcs, and their blocking property will depend on an existence of the activating subsets described previously. Only if such a subset exists, an inhibitor arc prevents an activation of the corresponding transition. An example of such behavior is given in part d) of Figure \ref{fig:4}. In that example, the inhibitor arc (painted red) requires 5 tokens. Six are available in place $p_0$ and five of them are mature enough (with a lifetime equal to $\gamma^{L}_{p_0}$ or greater) to form activating subset. In that case, however, this subset can be rather called a blocking one, because it allows the inhibitor arc to block the activation of transition $t_0$. Without the inhibitor arc, it could be active, because there are three old enough tokens in $p_1$.

\section{Conclusions}\label{sec5}

The proposed extended time Petri net has many very interesting properties, and a more detailed description of some of them is beyond the scope of this paper. Two features in particular, in our opinion, suggest that the proposal of a new time net combining the features of already known ones can be a good and interesting idea. First of all, it is often the case that when trying to create a model of a certain system, the net that best fits the format of the available data on the system is selected. Often, however, it is necessary to match data and net elements to each other because the specific net used perhaps well, but not ideally, fits the type of data that are available. Transitions in a TPN remain active for a certain period of time, followed by a moment of activation. Then, at the same moment, tokens are taken from the input places, while new tokens are created at the output places. In DPN/ITPN nets, an active transition is immediately activated, that is, it takes tokens, but the production of new ones requires some time. The difference is that in a TPN, in case of a conflict between many transitions sharing common places, tokens from a given place can simultaneously participate in the activation of many transitions until they are taken. In DPN/ITPN they cannot, since they are consumed immediately. If the system studied has elements that require either one or the other behavior, sometimes some conversions or simplifications must be performed inside the model. This, in turn, can further obscure the overall picture of the model. In the proposed extended time Petri net, a transition implicitly has both features, i.e., a time until activation and a separate production time counted when tokens have already been taken. Transformation of net elements is possible and can be done by modifying only the time data describing the transitions. Thus, there is no need to add some artificial elements to the net while transforming it to reflect either behavior.

The second important feature of the new net is that transitions and places are described by two and one time intervals, respectively, rather than by a specific single value of, for example, production time, as in a DPN. One can imagine many systems in which the time of execution of certain activities cannot be reduced to one specific and deterministic value, but it can at most be given as an approximation. Biological systems in particular are often described in this way because it is almost impossible to give the precise time of occurrence of, for example, a single chemical reaction. Thus, the proposed net turns out to be a very flexible tool that can be easily adapted to the data and knowledge of the system under study.   

\section*{Acknowledgements}
This work was partially supported by statutory funds of Poznan University of Technology.

%\section*{Declarations}
%\begin{itemize}
%\item Conflict of interest/Competing interests
%The authors declare that they have no conflict of interest.
%\end{itemize}

\bibliographystyle{abbrv}
\bibliography{myBibliography}

\end{document}